\documentclass[12pt]{article}

\usepackage[T1]{fontenc}
\usepackage[utf8]{inputenc}
\usepackage{tgpagella}

\usepackage[authoryear]{natbib}
\usepackage{amssymb}
\usepackage{amsfonts}
\usepackage{amsmath}
\usepackage{dsfont}
\usepackage{soul}
\usepackage[nohead]{geometry}
\usepackage{graphicx}
\usepackage{amsthm}
\usepackage{color}
\usepackage{comment}
\usepackage{setspace}
\usepackage{framed}
\usepackage{enumitem}
\usepackage[disable]{todonotes}
\usepackage{tikz}
\usepackage{hyperref}
\usepackage{rotating}
\usepackage{subfigure}
\usepackage[flushleft]{threeparttable}
\usepackage{multirow}
\newcommand{\cref}[2][1]{{\textup{(\hyperref[#2]{\ref*{#2}$_{#1}$})}}}
\usepackage{xr}
\usepackage{bm}

7

\newcommand{\eq}[1]{\begin{align}#1\end{align}}
\newcommand{\eqs}[1]{\begin{align*}#1\end{align*}}
\newcommand{\lt}{\left}
\newcommand{\rt}{\right}

\newcommand{\eps}[0]{\ensuremath{\varepsilon}}

\newcommand{\what}{\widehat}
\newcommand{\ap}{\alpha}
\newcommand{\bt}{\beta}
\newcommand{\ld}{\lambda}

\newcommand{\gm}{\gamma}

\newcommand{\beq}{\begin{eqnarray*}}
\newcommand{\eeq}{\end{eqnarray*}}

\numberwithin{equation}{section}
\theoremstyle{definition}

\makeatletter
\def\@biblabel#1{\hspace*{-\labelsep}}
\makeatother
\DeclareMathOperator*{\argmin}{argmin}

\begin{document}

\title{Desperate times call for desperate measures: government spending multipliers in hard times\thanks{We would like to thank the Seoul National University Research Grant in 2020, the Social Sciences and Humanities Research Council of Canada (SSHRC-435-2018-0275), the European Research Council for financial support (ERC-2014-CoG-646917-ROMIA) and the UK Economic and Social Research Council for research grant (ES/P008909/1) to the CeMMAP.}}
\date{\today}

\author{
Sokbae  Lee\thanks{%
Lee: Professor, Department of Economics, Columbia University, 420 West 118th Street,  New York, NY 10027, USA. E-mail: \texttt{sl3841@columbia.edu}.}   \and
Yuan Liao\thanks{
Liao: Associate Professor, Department of Economics, Rutgers University, 75 Hamilton St., New Brunswick, NJ 08901, USA. Email:
\texttt{yuan.liao@rutgers.edu}.}
\and Myung Hwan Seo\thanks{%
Seo: Associate Professor, Department of Economics, Seoul National University, 1 Gwanak-ro, Gwanak-gu, Seoul 08826, Korea. E-mail:
\texttt{myunghseo@snu.ac.kr}.}   \and
Youngki  Shin\thanks{%
Shin: Associate Professor, Department of Economics, McMaster University, 1280 Main St.\ W.,\ Hamilton, ON L8S 4L8, Canada. Email:
\texttt{shiny11@mcmaster.ca}.}
}

\maketitle

\begin{abstract}
\noindent
We investigate state-dependent effects of fiscal multipliers  and allow for endogenous sample splitting to determine whether the US economy is in a slack state. When the endogenized slack state is estimated as the period of the unemployment rate higher than about 12 percent, the estimated cumulative multipliers are significantly larger during slack periods than non-slack periods and are above unity. We also examine the possibility of time-varying regimes of slackness and find that our empirical results are robust under a
more flexible framework. Our estimation results point out the importance of  the heterogenous effects of fiscal policy and  shed light on the prospect of fiscal policy in response to economic shocks from the current COVID-19 pandemic. 
\end{abstract}

\thispagestyle{empty}



\onehalfspacing

\newpage
\setcounter{page}{1}
\pagenumbering{arabic}

\section{Introduction}

The debate over the role of fiscal policy during a recession has recently taken center stage again in macroeconomics. One particular topic that has received substantial attention is whether the multiplier effect of government spending is state-dependent. On the one hand,
in a series of papers, \cite{Auerbach12, Auerbach13a, Auerbach13b} used data from the USA as well as from OECD countries and provided empirical evidence supporting that the fiscal multiplier might be larger during recessions than expansions.
On the other hand,
\cite{RZ-2017} constructed new quarterly historical US data and reported that their estimates of the fiscal multipliers were below unity irrespective of the state of the economy.

In this paper, we contribute to this debate by estimating a threshold regression model that determines the states of the economy endogenously.
\cite{Auerbach12} estimated
smooth regime-switching  models using a seven quarter moving average of the output growth rate as the threshold variable.
Their primary results relied on a fixed level of intensity of regime switching. Instead of estimating the level of intensity jointly with other parameters in their model, they calibrated the level of intensity so that  the US economy spends about 20 percent of time in a recessionary regime.
In \cite{RZ-2017}, the baseline results assume that the US economy is in  a slack state if the unemployment rate is above 6.5 percent.
To check the baseline results,  \cite{RZ-2017} conducted various robustness checks using different thresholds.

To be consistent with the empirical literature, we build on  \cite{RZ-2017}: we use their dataset  and follow their methodology closely. Our main departure from the recent empirical literature is that we split the sample in a data-dependent way so that the choice of threshold level is determined endogenously.
It turns out that the endogenized threshold level of the unemployment rate is estimated at 11.97 percent, which is much higher than 6.5 percent adopted in \cite{RZ-2017}.
Using this new threshold level combined with the same data and specifications as in \cite{RZ-2017}, we find that the estimated fiscal multipliers are significantly different between the two states and above  unity for the high unemployment state.
Specifically, if the threshold level is 6.5 percent, the estimates of  two-year integral multipliers  are around 0.6 regardless of the state of the economy. However, if the threshold level is 11.97 percent, the estimates are 1.58 for the high employment state and 0.55 for the low employment state, respectively.
If we look at observations used in estimation, there is no period after World War II with the unemployment rate higher than 11.97 percent. In fact, there is only one timespan of severe slack periods in 1930s. In other words, the period of the Great Depression is isolated from other periods, as an outcome of our estimation procedure.
Therefore, our estimation results suggest that (i) the fiscal multiplier can be larger than unity if the slackness of the economy is very severe and that
(ii) the post World War II period does not include the severe slack state and thus, our estimates for the high unemployment state are not applicable to moderate recessions in the post WWII period.
However, after the outbreak of the COVID-19 pandemic,  the US unemployment rate  rose to 14.7\% in April 2020.\footnote{Source: US Bureau of  Labor Statistics, \url{https://www.bls.gov/news.release/empsit.nr0.htm}, accessed on May 25, 2020.} 
Therefore, the estimation results in this paper shed light on the prospect of the fiscal policy in response to the current economic shocks.
We also examine the possibility of time-varying regimes of slackness
by including a time dummy for the post WWII period
and find that our empirical results are robust under this more flexible framework.
All the computer codes and data files for replication are available at \url{https://github.com/yshin12/llss-rz}.

The remainder of the paper is organized as follows. In Section \ref{model}, we describe the econometric model and present empirical results.
In Section \ref{sec:conclusions}, we give concluding remarks.

\section{Model and Empirical Results}\label{model}

In this section, we give a brief description of the methodology developed by \citet[RZ hereafter]{RZ-2017}.
They consider the state-dependent local projection method of \citet{jorda2005estimation}. Their baseline regression model for each horizon $h$ has the following form  (see equation (2) in RZ):
\begin{align}\label{lp-eq}
\begin{split}
x_{t+h} = & I_{t-1} \lt(\ap_{A,h} + \psi_{A,h}(L)z_{t-1}+\bt_{A,h} shock_t \rt) \\
& \hskip10pt +  (1-I_{t-1}) \lt(\ap_{B,h} + \psi_{B,h}(L)z_{t-1}+\bt_{B,h} shock_t \rt) +\eps_{t+h},
\end{split}
\end{align}
where $I_{t}(\cdot)$ is a dummy variable denoting the state of the economy, $x_t$ is the variable of interest, $z_t$ is a vector of control variables including  GDP, government spending, and lags of the defense news variable, $\psi(L)$ is a polynomial of order 4 in the lag operator, and $shock_t$ is the defense news variable.

Recall that RZ assume that the economy is in the slack state when the unemployment rate is above 6.5 percent. We instead adopt a threshold regression model and parameterize $I_t = 1\{unemp_t > \tau\}$, where $1\{\cdot\}$ is an indicator function and $unemp$ denotes the unemployment rate. In other words, we estimate the model that endogenously determines the slack states that fit the data best. Specifically, we estimate the following model using the least squares \citep[see, e.g.,][]{hansen2000sample, hidalgo2019robust}:
\begin{align}\label{ols-eq}
\begin{split}
GDP_t = & 1\{unemp_{t-1}>\tau\} \lt(\ap_{A} + \psi_{A}(L)z_{t-1}+\bt_{A} shock_t \rt) \\
& \hskip10pt +  1\{unemp_{t-1} \le \tau\}  \lt(\ap_{B} + \psi_{B}(L)z_{t-1}+\bt_{B} shock_t \rt) +\eps_{t}.
\end{split}
\end{align}
To estimate the threshold regression model in \eqref{ols-eq}, 
define the objective function
\begin{align*}
Q_T(\tau,\theta):= \sum_{t=1}^T
\Big[
GDP_t -  1\{unemp_{t-1}>\tau\} \lt(\ap_{A} + \psi_{A}(L)z_{t-1}+\bt_{A} shock_t \rt) \\
 \hskip10pt -  1\{unemp_{t-1} \le \tau\}  \lt(\ap_{B} + \psi_{B}(L)z_{t-1}+\bt_{B} shock_t \rt)
\Big]^2,
\end{align*}
where $\theta:=(\ap_A, \psi_A(L),\bt_A , \ap_B,\psi_B(L),\bt_B)$. Note that the model \eqref{ols-eq} is linear in $\theta$ conditional on $\tau$. Thus, we obtain the (restricted) OLS estimator $\what{\theta}(\gm)$ easily for any given $\gm$. Then, the threshold parameter $\gm$ can be estimated by minimizing the profiled objective function:
\eqs{
  \what{\tau} & := \argmin_{\tau \in \mathcal{T}} Q^*_T(\gm)
}where $Q^*_T(\gm):=Q_T\left(\gm,\what{\theta}(\gm)\right)$. To estimate this model, it is necessary to specify the parameter space $\mathcal{T} $ for $\tau$. We set it to be the interval between the 5 and 95 percentiles of the unemployment rates in the dataset and estimate $\widehat{\tau}$ by the grid search method.

In our view, the threshold regression model above provides a natural way to endogenize the level of slackness since there is a change point at $\tau$ for GDP in the model.
Note that the level of the slackness is determined endogenously by fitting the regression model for GDP in \eqref{ols-eq} and then it is imposed in the specification of $I_{t-1}$ in \eqref{lp-eq}.  Considering that both RZ and \cite{Auerbach12, Auerbach13a, Auerbach13b} determine the criterion for the economic slackness based on the researchers' discretion, it is novel to determine the threshold point endogenously. Furthermore, as we will see in the next section, the endogenous threshold estimate is beyond the range of the values that RZ considered for a robustness check.

In general, estimating the change point $\tau$ tends to be robust to model misspecification. Specifically, in our context, the local projection argument may imply that the model \eqref{ols-eq} is potentially misspecified; however, it is worthwhile to emphasize that the change-point estimation tends to be robust against mild misspecification in the regression function employed in each regime, as shown by e.g.\ 
\citet{Bai:2008}.

Before looking at the estimation results, we briefly describe the dataset adopted in our empirical analysis. RZ constructed new quarterly US data from 1889 to 2015 for their analysis. The main variables include real GDP, real government spending, the unemployment rate, and the defense news series. The real GDP data come from Historical Statistics of the United States for 1889--1928 and from the National Income and Product Accounts from 1929 to 2015. Real government spending is calculated by dividing all federal, state, and local purchases by the GDP deflator. The unemployment rates before 1948 were calculated by interpolating \citet{weir1992century}'s series and the NBER Macrohistory database. Finally, the defense news series is constructed by the narrative method of \citet{ramey2011identifying}, which measures changes in the expected present discounted value of government spending. For additional details of the dataset, we refer to \citet{RZ-2017}.

\subsection{Endogenous Sample Splitting}

\begin{figure}[htbp]
\begin{center}
\caption{Least Squares Objective Function}\label{fg:obj-values}
\vskip10pt
\begin{tabular}{cc}
Full sample & Subsample with $unemp < 11.97$ \\
\includegraphics[scale=0.4]{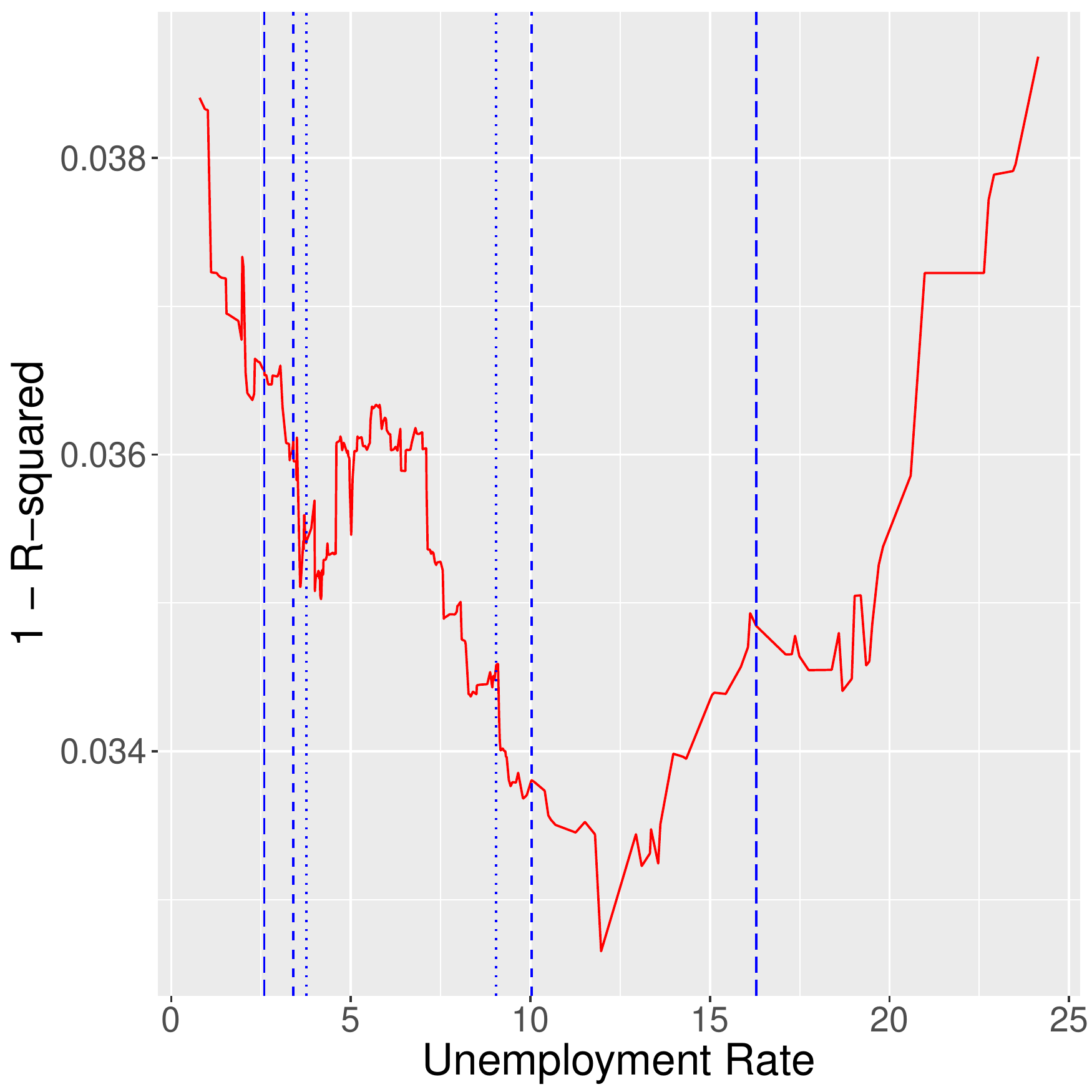} & \includegraphics[scale=0.4]{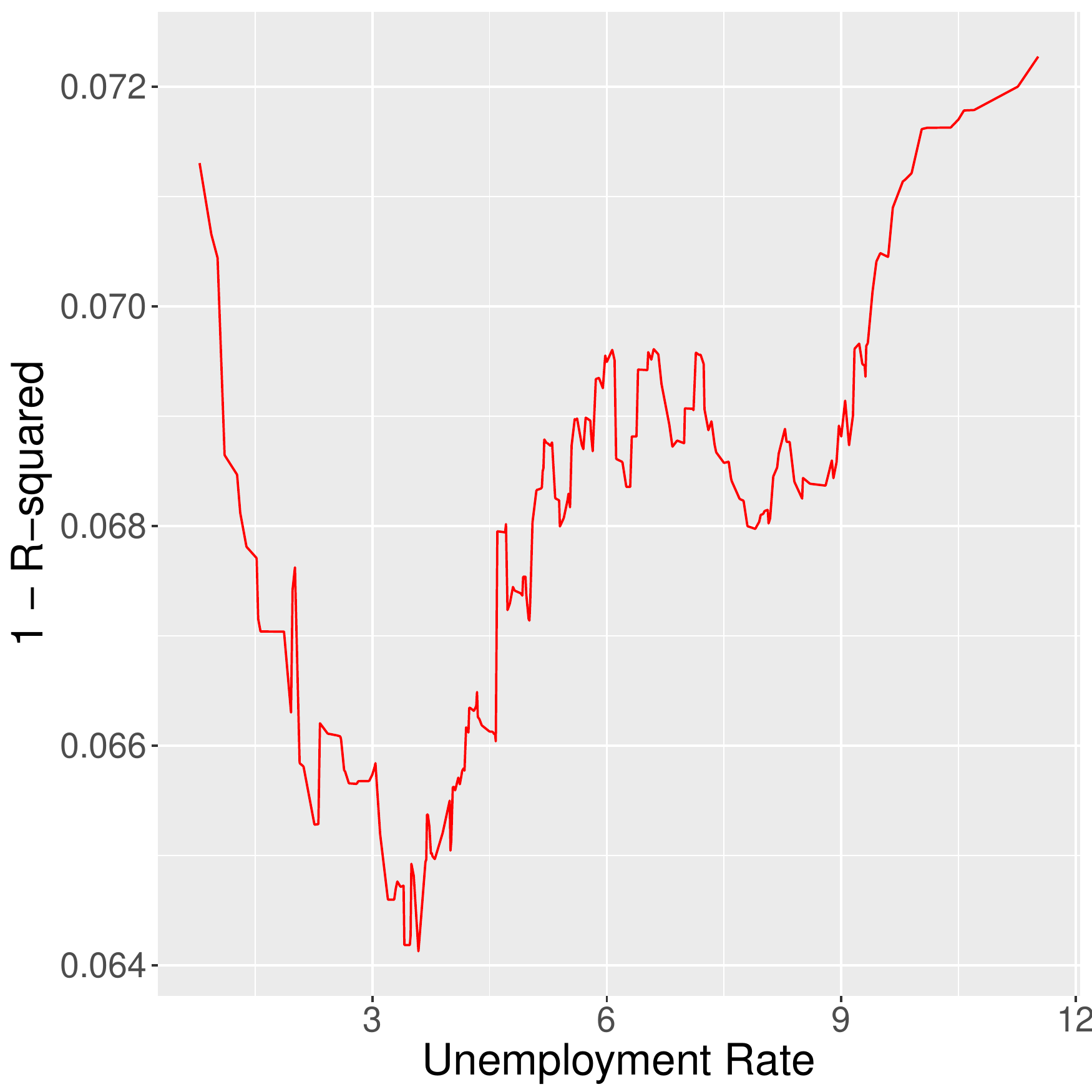}
\end{tabular}
\end{center}
\noindent \footnotesize{Note: In the left-hand panel, the long-dashed vertical lines are the 5  and 95 percentiles of the empirical distribution of the unemployment rate. The dashed vertical lines are the 10  and 90 percentiles and the dotted lines are the 15 and 85 percentiles, respectively. }
\end{figure}

Using the same dataset constructed by RZ, we obtain $\hat{\tau}=11.97\%$ for the threshold parameter. This estimate is even higher than 8 percent, which RZ used for their robustness check.
To appreciate our estimation result, we plot the profiled least squares objective function ($1-R^2$) as a function of $\tau$ in the left-panel of Figure \ref{fg:obj-values}.

It can be seen that the minimizer is well separated at 11.97\%, which gives the graphical verification of $\hat{\tau}$.
On the contrary, there is even no local minimum around RZ's threshold value at 6.5\%.
To check the possibility of the second threshold level below 11.97\%, we re-estimated the model with the subsample for which the unemployment rate is lower than 11.97\%. The right-hand panel indicates that there could be a second threshold around 4 percent, but  not around 6.5\%.

\begin{figure}[htbp]
\begin{center}
\caption{Inference for Multiple Regimes}\label{fg:irf-dummy3-1}
\begin{tabular}{cc}
Full sample & Subsample ($unemp < 11.97$) \\
\includegraphics[scale=0.4]{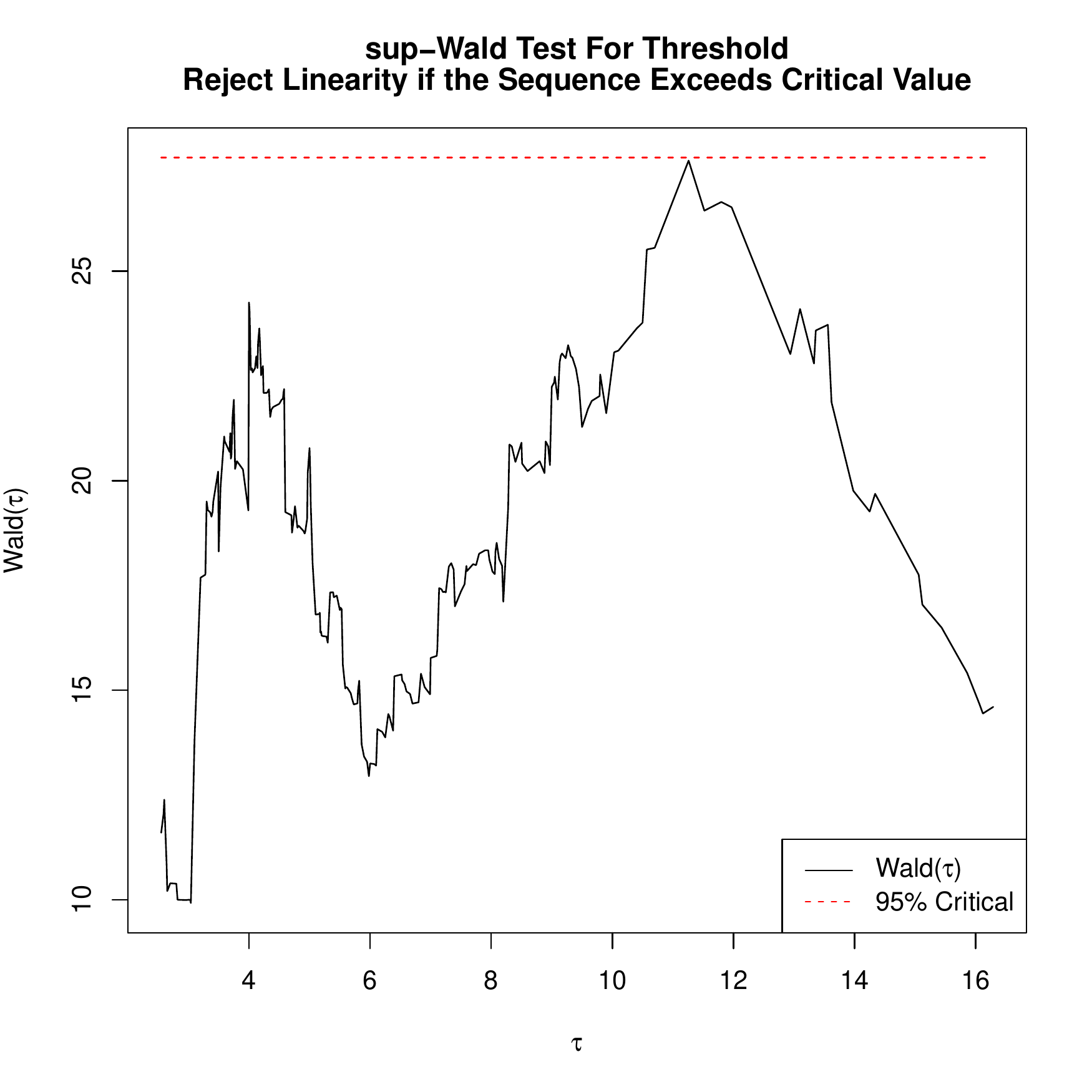} & \includegraphics[scale=0.4]{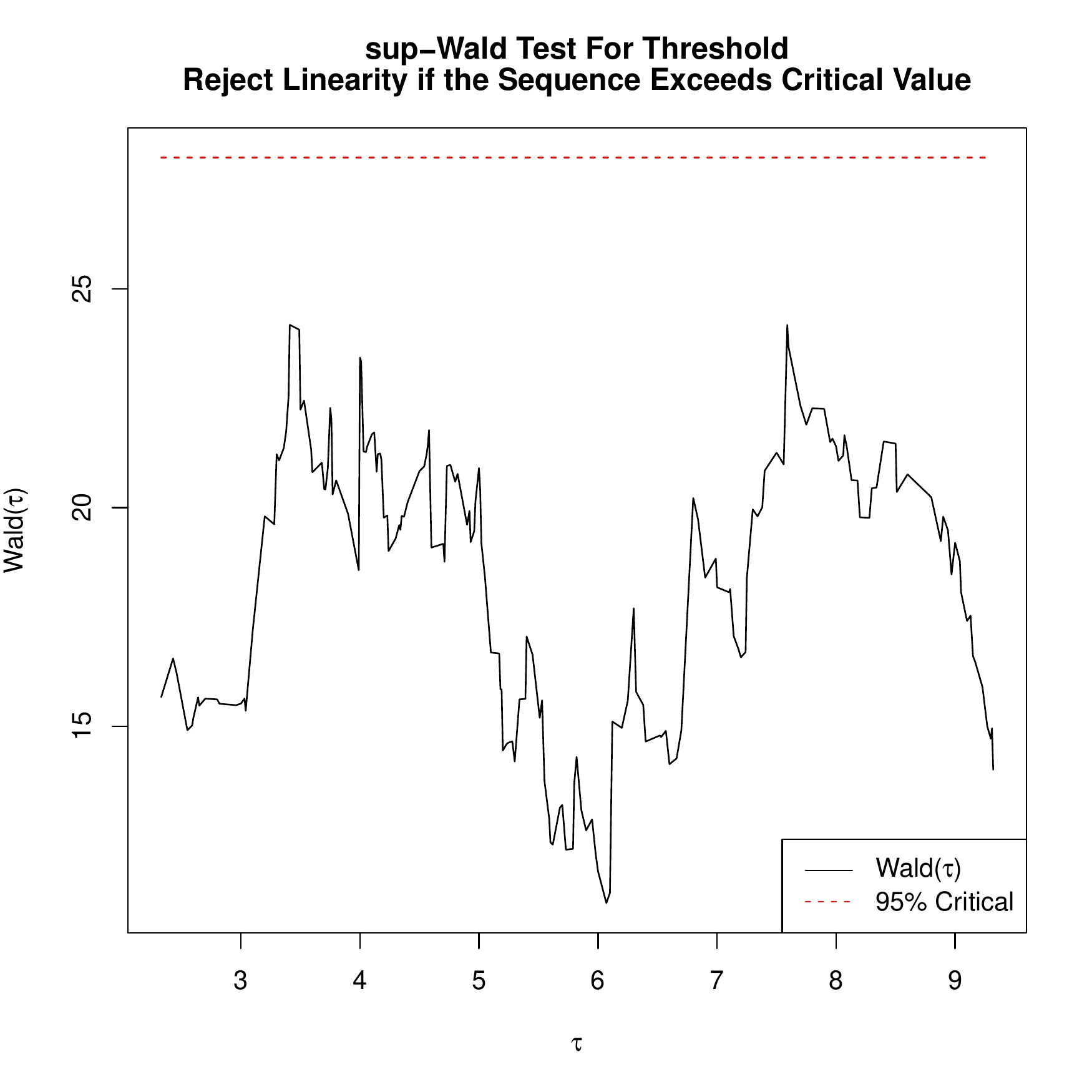}
\end{tabular}
\end{center}
\noindent \footnotesize{Note: The red dashes line denotes the 95\% critical value for the existence of the threshold point. 
In the left panel, we confirm that the Wald test statistic at $\tau=11.97$ is very close to the 95\% critical value. 
In the right panel, we use the subsample and test if there exists an additional threshold point. The result confirm that there is no additional threshold point in the subsample.
}
\end{figure}

We test for the existence of the threshold for the whole sample and for the subsample with $unemp < 11.97$ by adopting the sup-Wald test in \citet{Hansen:1996}. 
Figure \ref{fg:irf-dummy3-1} gives a graphical summary of the testing results.
We set the number of bootstraps to 2,000 and the trimming ratio to 5\%. We use the heteroskedasticity-robust test statistic. The bootstrap p-value for the whole sample is 0.053 and we can reject the null hypothesis of no threshold effect at the 10\% significance level. For the subsample with the unemployment rate below 11.97, the bootstrap p-value for the same test is 20.3\%. Thus, we conclude that there is  mild evidence for the single threshold in the data. Finally, the 95\% confidence interval for the threshold variable is $(11.97, 13.56)$.

\begin{figure}[htbp]
\begin{center}
\caption{Periods of Slack States over GDP and Unemployment}\label{fg:slack-states}
\vskip10pt
\begin{tabular}{cc}
GDP & Unemployment \\
\includegraphics[scale=0.4]{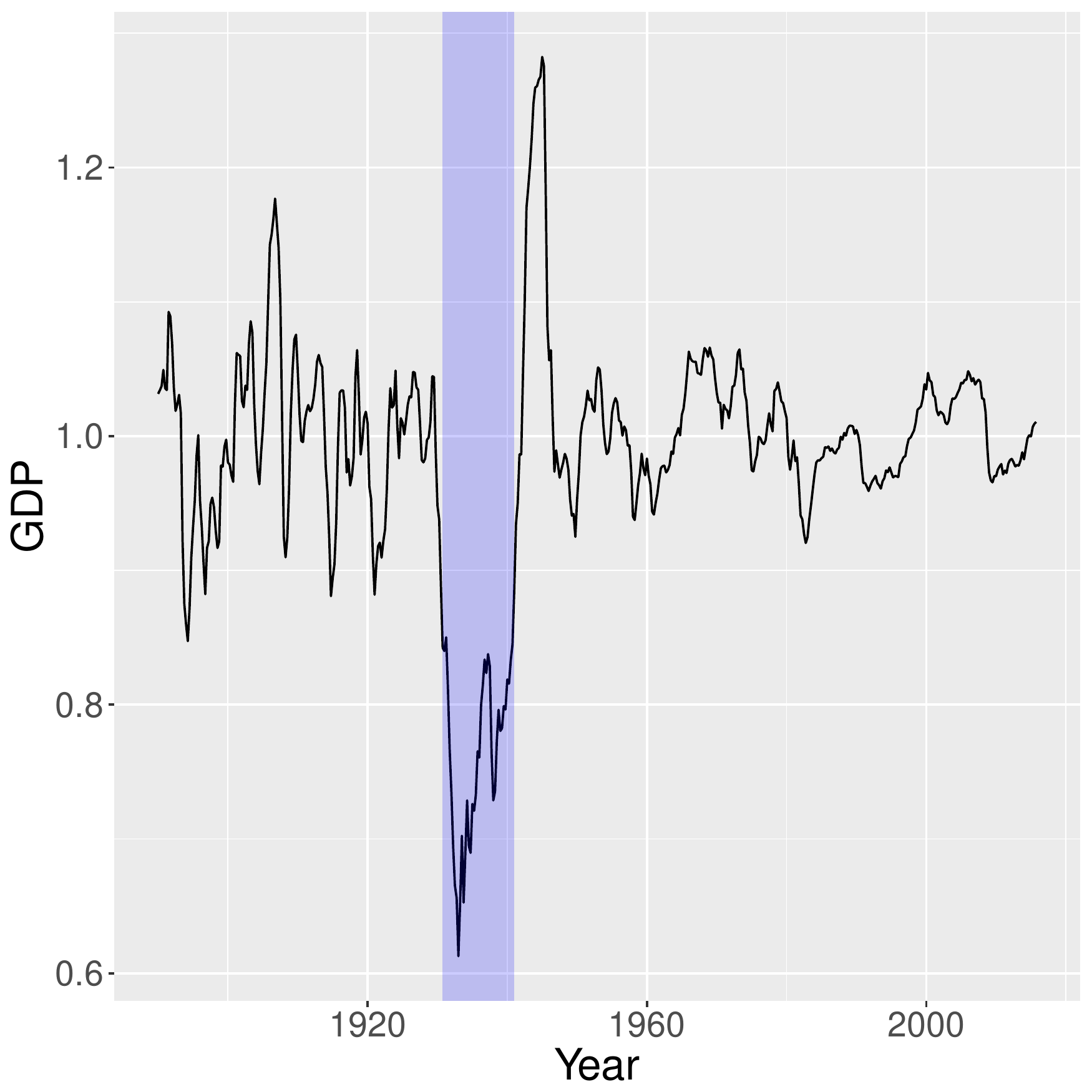} & \includegraphics[scale=0.4]{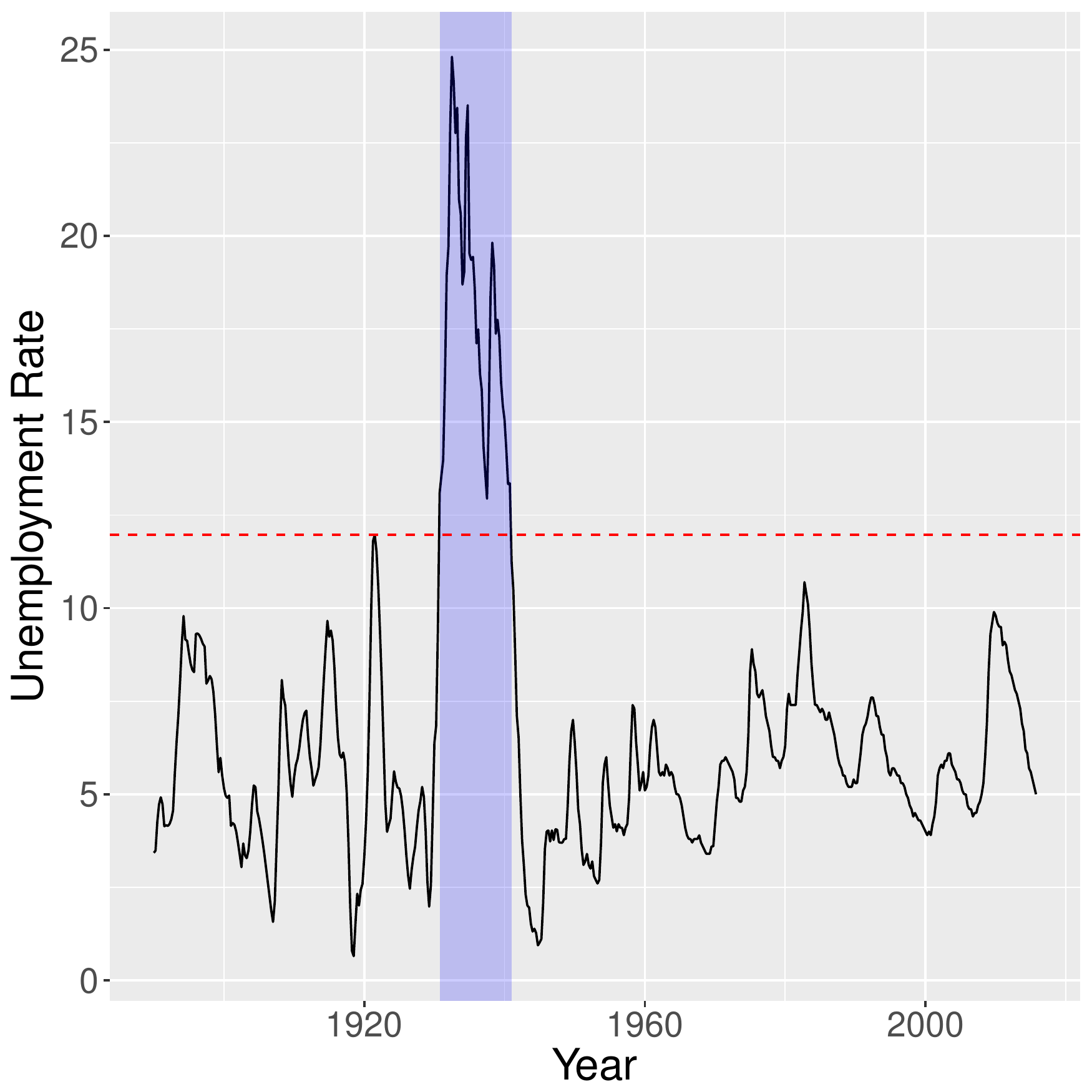}
\end{tabular}
\end{center}
\noindent \footnotesize{Note: $GDP$ denotes real per capita GDP divided by trend GDP. The red dashed line in the right panel is the change-point estimate, $\hat{\tau}=11.97$. The blue shaded area denotes the slack states estimated from the data.}
\end{figure}

The periods with high unemployment rates are relatively rare. The US economy spent less than 10 percent of time in the new slack regime defined by 11.97 percent.
The shaded areas in Figure \ref{fg:slack-states} show slack periods over GDP and unemployment rates. There is only one timespan of severe slack periods from 1930Q3 to 1940Q3, namely the Great Depression. We call this new slack periods as severe slack states (``hard times'') compared to moderate slack states in RZ. There is no period after WWII that belongs to the hard times in this dataset.  However, the current recession belongs to the hard times, as the unemployment rose to 14.7\% in April, 2020.

\subsection{State-Dependent Cumulative Multipliers}

\begin{figure}[hpt]
\begin{center}
\caption{Cumulative Multipliers}\label{fg:multiplier}
\vskip10pt
\begin{tabular}{c}
\underline{LLSS: Threshold: 11.97\%} \\
\includegraphics[scale=0.4]{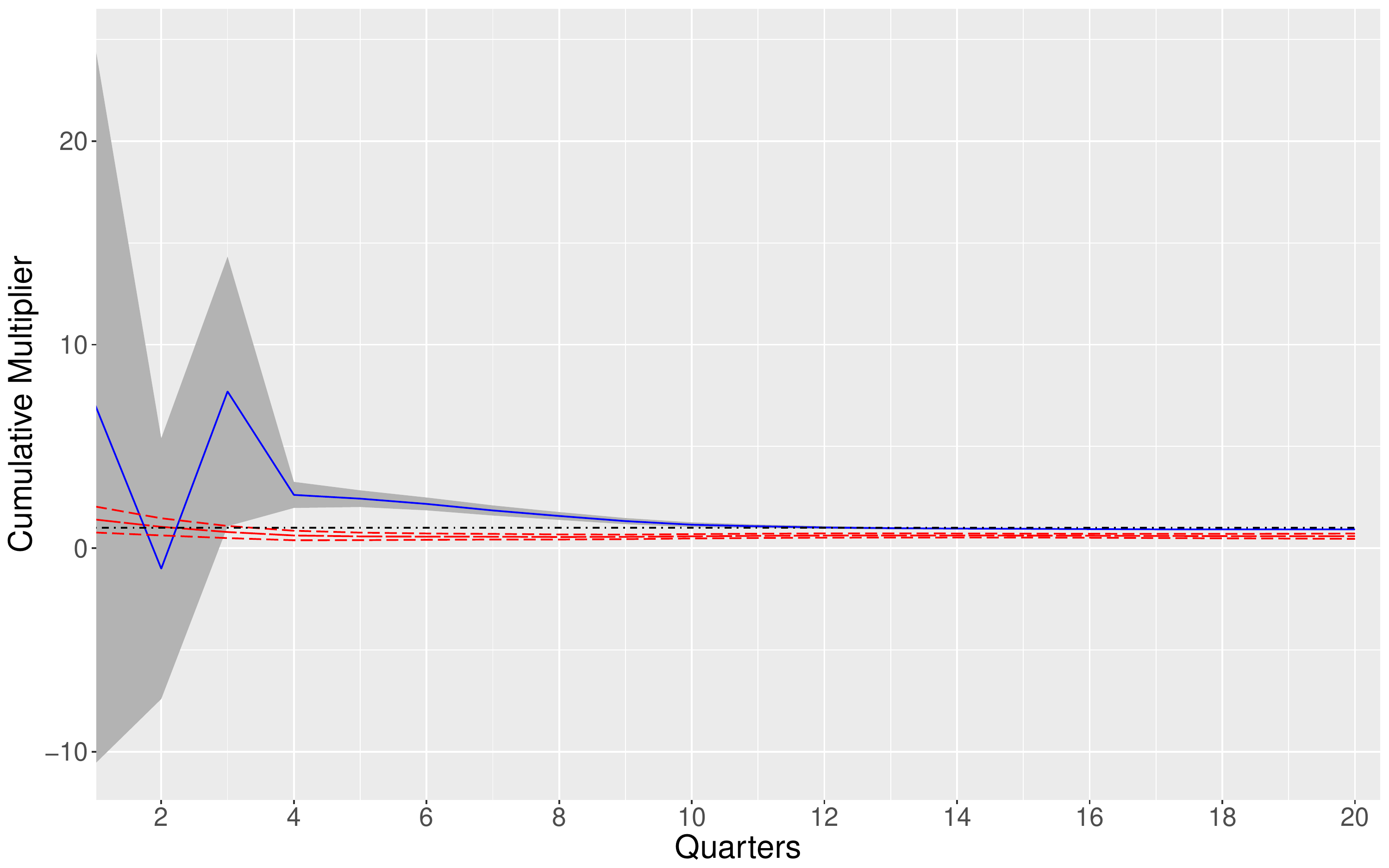} \\
\\
\underline{RZ: Threshold: 6.5\%}  \\
\includegraphics[scale=0.4]{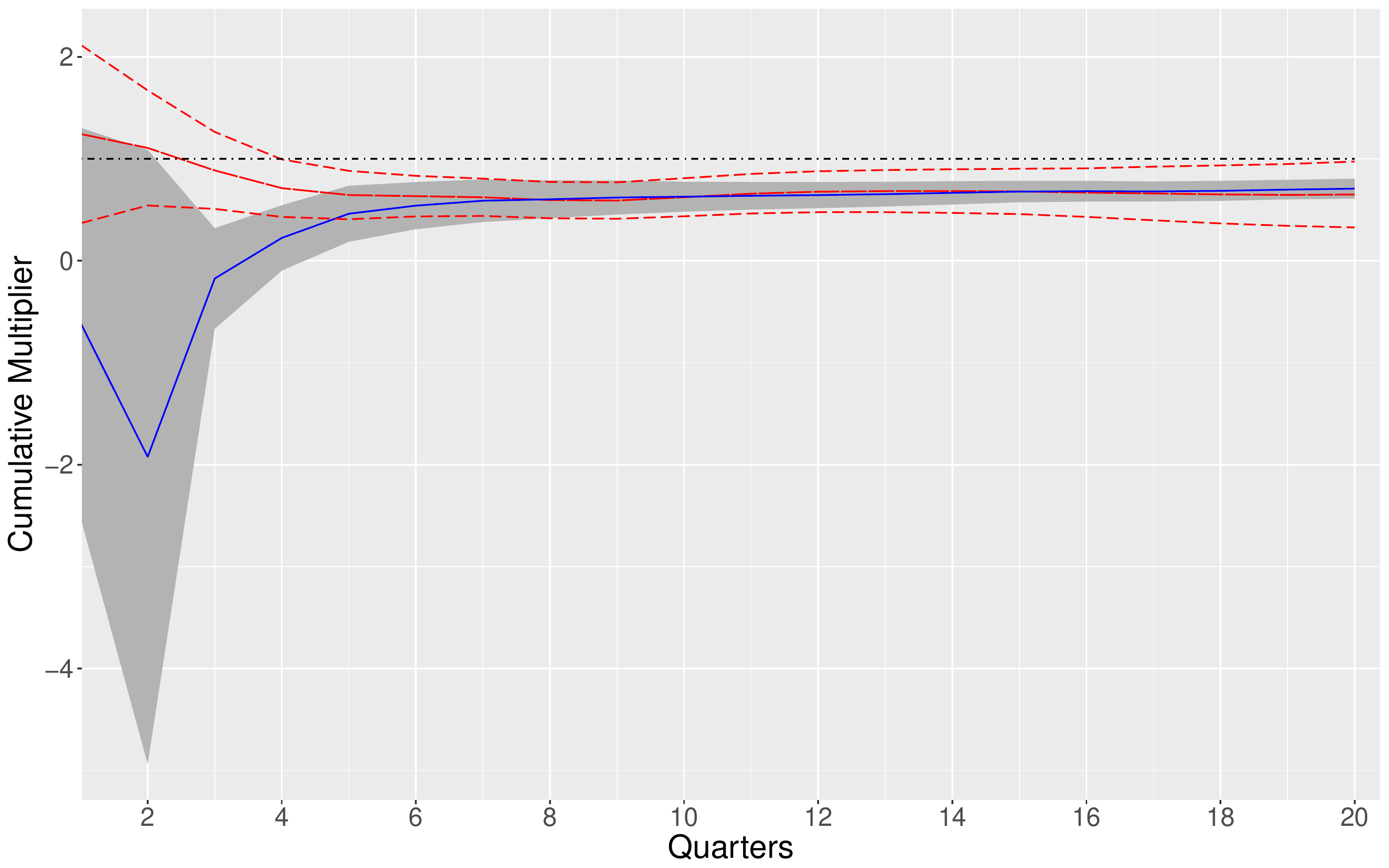}
\end{tabular}
\end{center}
\noindent \footnotesize{Note: The blue solid line denotes cumulative multipliers for slack states (high unemployment) and the red dashed line for non-slack states (low unemployment). The 95\% pointwise confidence bands are also presented along with cumulative multipliers. We also draw a dot-dashed horizontal line at multiplier=1.}
\end{figure}

We now report the estimation results of  the cumulative multipliers under endogenous sample splitting.
It turns out that the new regime classification
 produces quite different implications. Following RZ, we adopt the local projection method in \citet{jorda2005estimation} and use the military news as an instrument. Figure \ref{fg:multiplier} reports the cumulative multiplier over 5 years (20 quarters) in each regime. To make the comparison straightforward, we also show the estimation results of \citet{RZ-2017} next to our results.

\begin{figure}[htbp]
\begin{center}
\caption{Government Spending and GDP Responses to News Shock}\label{fg:ifs}
\vskip10pt
\begin{tabular}{cc}
Government Spending & GDP \\
\includegraphics[scale=0.45]{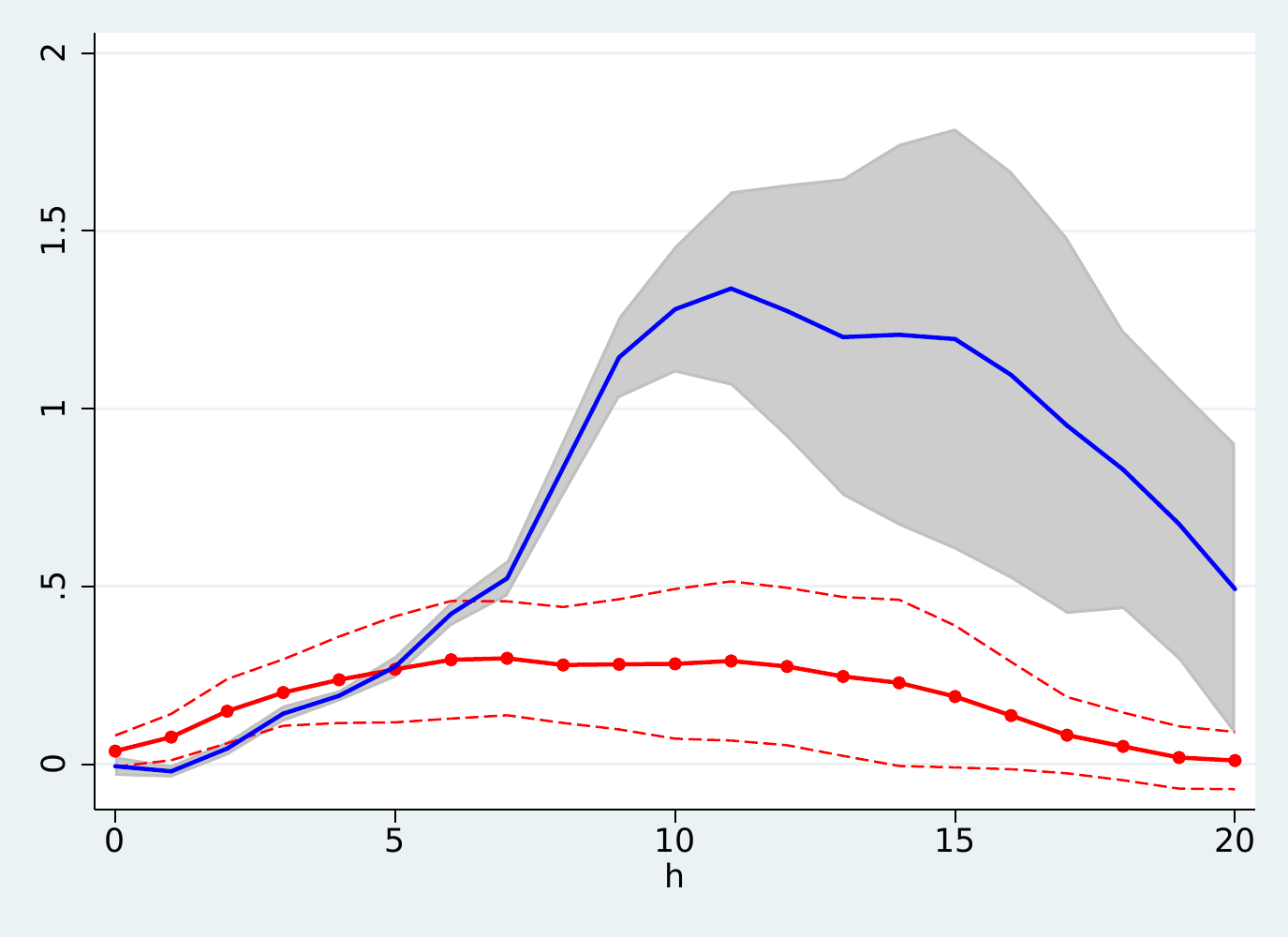} & \includegraphics[scale=0.45]{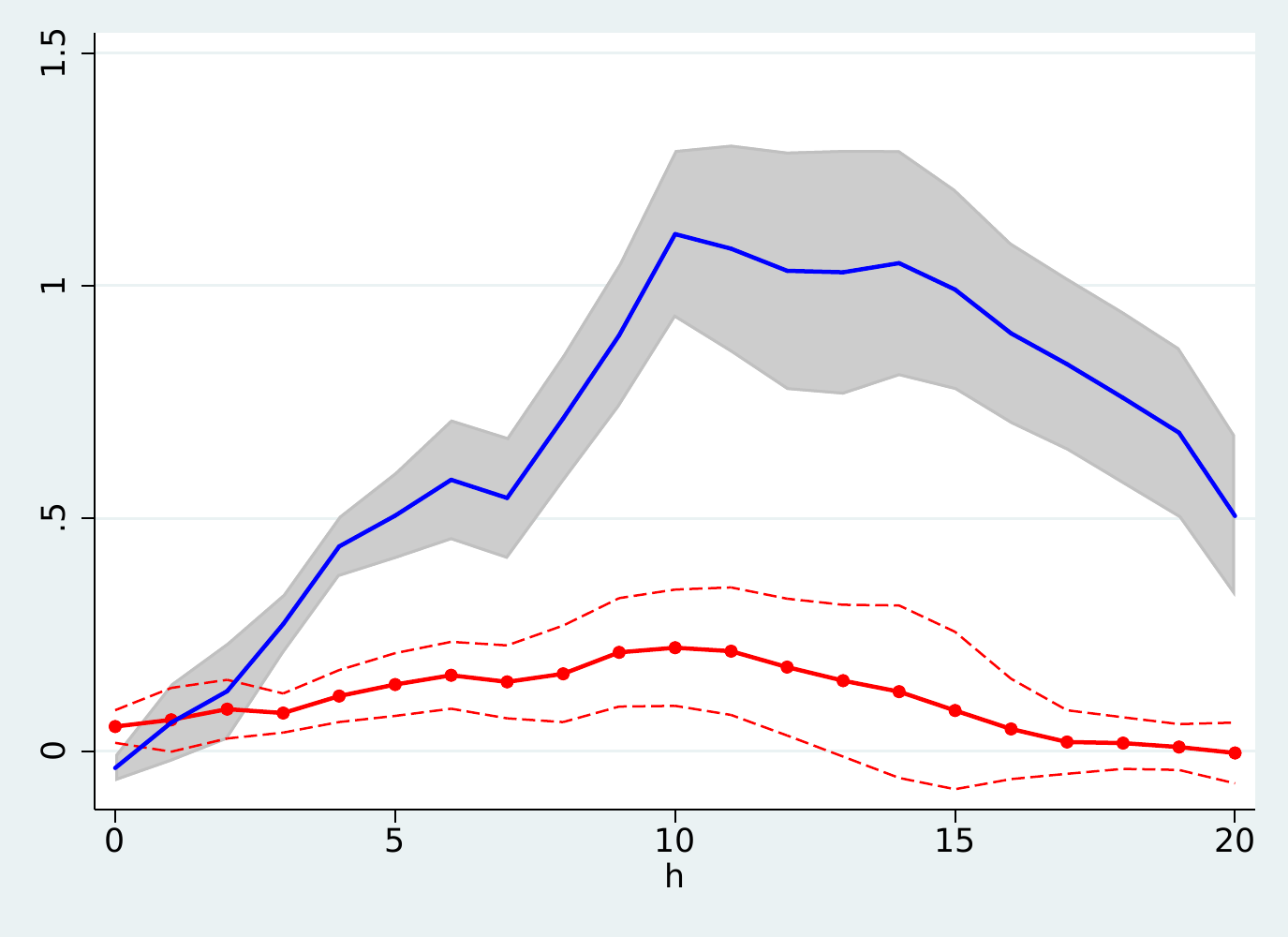}
\end{tabular}
\end{center}
\noindent \footnotesize{Note: A news shock is equal to 1 percent of GDP. The red line with circles denotes the impulse response function in non-slack periods and the blue solid line denotes the same function in slack periods. The related 95\% pointwise confidence bands are also provided. The threshold point dividing slack/non-slack periods is  $\hat{\tau}=11.97$ estimated from the data.}
\end{figure}

When the 6.5\% threshold is used in classification of slack state (that is, the moderate slack state),
the multipliers in the high-unemployment state are negative up to 3 quarters and are indistinguishable to those in the low-unemployment state after 6 quarters. It is counterintuitive to observe that the multipliers are higher for the low unemployment state. On the other hand, if the 11.97\% threshold is adopted (that is, the severe slack state),
the multipliers in the high-unemployment state are mostly positive and largely above those in the low-unemployment state
and are around unity after 10 quarters.
In other words,
the multipliers are all less than unity in the case of  the moderate slack state; however, they are substantially higher in the case of  the severe slack state.  These results are robust to the choice of the instrumental variable.
As additional empirical results, Figure \ref{fg:ifs} depicts  the impulse response functions in non-slack and slack periods, respectively.
Both government spending and GDP responses are much higher in slack periods.

\begin{table}[htbp]
\caption{Estimates of Cumulative Multipliers}\label{tb2:multiplier}
\centering
\begin{threeparttable}
\begin{tabular}{cccc}
\\
\hline
    & High             & Low                  & P-value for difference\\
       & Unemployment     & Unemployment        &  in multipliers \\

\hline
\\
\multicolumn{4}{l}{\underline{Panel A:  Threshold at 11.97\%}} \\
\multicolumn{4}{l}{\textbf{Military News Shock}} \\
2 year integral      & 1.58         & 0.55           & 0.000\\
                  & (0.099)     & (0.064)         & \\
4 year integral      & 0.94        & 0.61           & 0.000 \\
                  & (0.017)     & (0.050)         & \\
\\
\multicolumn{4}{l}{\textbf{Blanchard-Perotti Shock} } \\
2 year integral      & 1.65         & 0.34           & 0.005\\
                  & (0.425)     & (0.105)         & \\
4 year integral      & 1.23        & 0.40           & 0.000 \\
                  & (0.130)     & (0.104)         & \\
\\
\multicolumn{4}{l}{\textbf{Combined} } \\
2 year integral      & 2.21         & 0.35           & 0.000\\
                  & (0.406)     & (0.092)         & \\
4 year integral      & 1.11       & 0.46           & 0.000 \\
                  & (0.108)     & (0.086)         & \\

\hline \\
\multicolumn{4}{l}{\underline{Panel B. Threshold at 6.5\%}} \\
\multicolumn{4}{l}{\textbf{Military News Shock}} \\
2 year integral      & 0.60         & 0.59             & 0.954 \\
                  & (0.095)        & (0.091)       &     \\
4 year integral      & 0.68        & 0.67          & 0.924    \\
                  & (0.052)     & (0.121)       &      \\
\\
\multicolumn{4}{l}{\textbf{Blanchard-Perotti Shock} } \\
2 year integral      & 0.68         & 0.30             & 0.005 \\
                  & (0.102)        & (0.111)       &     \\
4 year integral      & 0.77        & 0.35          & 0.001    \\
                  & (0.075)     & (0.107)       &      \\
\\
\multicolumn{4}{l}{\textbf{Combined} } \\
2 year integral      & 0.62         & 0.33             & 0.099 \\
                  & (0.098)        & (0.110)       &     \\
4 year integral      & 0.68        & 0.39          & 0.021    \\
                  & (0.052)     & (0.110)       &      \\
\\
\hline
\end{tabular}
\begin{tablenotes}
\small
\item Note: The p-values for difference in multipliers are calculated by the HAC-robust p-values in \citet{NW87}.  Panel A is based on our threshold estimate (11.97\%). Panel B comes from \citet{RZ-2017} where the threshold point (6.5\%) is chosen by the authors.
\end{tablenotes}
\end{threeparttable}
\end{table}

 In Table \ref{tb2:multiplier}, we report the 2-year and 4-year cumulative multipliers when we use the military news, \citet{BP:02} shock, and the combined variable of these two as an instrument, respectively.  The basic implication does not change. The estimates of the 2-year multiplier vary from 1.58 to 2.21 and the 4-year multipliers are around 1.
The main implication from our empirical results is that
fiscal multipliers can be significantly larger during severe recessions than in normal periods.

We illustrate the difference between our results and those in RZ by comparing the effects of the COVID-19 stimulus package. The COVID-19 pandemic and the following economic lockdown increased the US unemployment rate up to 14.7 percent in April 2020. This is the highest unemployment rate since World War II. To mitigate the economic hardship, the US congress has passed the COVID-19 stimulus package (the CARES act) whose total amount is 2 trillion dollars. In Table \ref{tb:covid-19}, we report the difference of the estimated multi-year integral effects of the stimulus package when we use the multipliers in this paper and those in RZ. We assume that 25\% of the total amount (500 billion dollars) will be spent in the immediate quarter and use the cumulative multiplier estimates based on the military news shock. Two approaches provide quite different results of the policy effect. Over two years, the difference between the two estimates is 490 billion dollars. The gap decreases over time but it is still 70 billion dollars after 5 years. Therefore, we conclude that the endogenous threshold estimate gives quite different results of the fiscal policy effect, especially when the slackness of the economy is severe.
\begin{table}[]
\caption{GDP Increases Caused by the COVID-19 Stimulus Package (in \$ bn)}\label{tb:covid-19}
\centering
\begin{threeparttable}
\begin{tabular}{l ccc}
\\
\hline
    & LLSS                       & RZ                          & \multirow{2}{*}{Difference}  \\
    & (Threshold at 11.97\%)     & (Threshold at 6.5\%)        &  \\

\hline
2 year integral      & 790        & 300           & 490 \\
3 year integral      & 510        & 355           & 155 \\
4 year integral      & 470        & 340           & 130 \\
5 year integral      & 465        & 395           & 70 \\
\hline
\end{tabular}
\begin{tablenotes}
\small
\item Note: The estimates denote the increased cumulate GDP when the US government spends 500 billion dollars in the period of high unemployment (14.7\%). Military news shocks are used as an instrument.
\end{tablenotes}
\end{threeparttable}
\end{table}

\subsection{Possibly Time-Varying Regimes}
In this subsection, we explore the possibility of time-varying regimes of slackness.
One might be worried that the US economy changed after WWII such that the level of slackness changed from the pre WWII period to the post WWII period. To deal with this issue, we extend the endogenous sample splitting to the following specification:
\eqs{
I_{t-1} = 1\{unemp_{t-1} + \tau_1 d_{t-1}-\tau_0 > 0 \},
}
where $d_t=1$ if $t$ is greater than or equal to 1945Q4.
The resulting regression model has the following form:
\eqs{
GDP_t = & 1\{unemp_{t-1} + \tau_1 d_{t-1}-\tau_0 > 0 \} \lt(\ap_{A} + \psi_{A}(L)z_{t-1}+\bt_{A} shock_t \rt) \\
& \hskip10pt +  1\{unemp_{t-1} + \tau_1 d_{t-1}-\tau_0 \leq 0 \}  \lt(\ap_{B} + \psi_{B}(L)z_{t-1}+\bt_{B} shock_t \rt) +\eps_{t}.
}
To estimate this model,  we need to optimize the least squares objective function with respect to
unknown parameters 
jointly.
The parameters could be estimated through the profiling method as explained in Section 2. Specifically, one may first estimate the slope parameters $\theta:=(\theta_A,\theta_B)=( \alpha_A, \psi_{A}, \beta_{A}, \alpha_B, \psi_{B}, \beta_{B})$ given $\tau:=(\tau_0, \tau_1)$ and then optimize the profiled objective function over $\tau$ by the 2-dimensional grid search.

We adopt more efficient computational algorithms developed in our previous work \citep{lee2018factor} with the aid of mixed integer optimization (MIO). To explain the algorithm, we first define some notation: $y_t:=DGP_t$, $f_t:=(unemp_{t-1}, d_{t-1}, -1)$, and $x_t:=(1, z_{t-1}, shock_t)$. Then, the least squares estimator can be written as
\begin{align}
  (\what{\tau}, \what{\theta}_B, \what{\delta}) := \argmin_{\tau, \theta_B, \delta} \sum_{t=1}^T \left[y_t - x_t'\theta_B - x_t'\delta 1\{f_t'\tau>0 \}\right]^2
\end{align}
where $\delta = \theta_A-\theta_B$. Instead of multi-dimensional grid search over $\tau$, \citet{lee2018factor} propose an equivalent optimization problem by introducing a set of binary parameters $d_t:=1\{f_t'\gm > 0\}$ and $\ell_{j,t} = \delta_j d_t$ for $j=1,\ldots, d_x$, where $d_x$ is the dimension of $x_t$. The new objective function can be written as
\eq{
  \sum_{t=1}^T \left[ y_t - x_t \theta_B - \sum_{j=1}^{d_x} x_{j,t} \ell_{j,t} \right]^2.
}The equivalent optimization problem becomes a mixed integer programming problem with some additional constraints. The new optimization problem can be solved efficiently by the modern MIO solvers such CPLEX and GUROBI. One can solve the optimization jointly or by iterating between $(\theta_B,\delta)$ and the remaining parameters. The advantage of the new algorithm is that one can construct and estimate the model, where the regimes are determined in a more sophisticated way by a multi-dimensional factor $f_t$. We refer to \citet{lee2018factor} for additional details.

By applying the joint and iterative algorithms proposed in that paper, we obtain the following results:
\eqs{
\mbox{Joint algorithm: }(\hat{\tau}_1, \hat{\tau}_0) & = (-1.82, 11.97),~obj = 0.0002636456,\\
\mbox{Iterative algorithm: }(\hat{\tau}_1, \hat{\tau}_0) & = (0.56, 11.97),~ obj = 0.0002636456.
}
That is, two algorithms yield different estimates but the same objective function values.
It turns out that the regimes determined by two estimates are identical; that is, $\hat{\tau}_1$ has no role in determining slack periods.

In addition, we apply the model selection algorithm  proposed in our previous work \citep{lee2018factor}.
Specifically, we specify the penalized least squares objective function with the penalty term consisting of a tuning parameter $\lambda > 0$ times the number of non-zero coefficients. The resulting  specification of the endogenous sample splitting rule is as follows:
\eqs{
  \sum_{t=1}^T \left[ y_t - x_t \theta_B - \sum_{j=1}^{d_x} x_{j,t} \ell_{j,t} \right]^2 + \ld |\tau|_0,
}
where $|\cdot|_0$ is an $\ell_0$ norm of a vector. We implement it using MIO with  $\ld=\hat{\sigma}^2\log(T)/T$,
 where $T$ is the sample size and $\hat{\sigma}^2=0.00027$ is estimated from the baseline model with a single threshold at 11.97\%.
When we apply the penalized estimation algorithm, we find that the $\tau_1$ estimate becomes zero and is dropped from the model.
Therefore, there is no empirical evidence that supports time-varying regimes of slackness.

\section{Conclusions}\label{sec:conclusions}

We have investigated state-dependent effects of fiscal multipliers and have found that it is crucial how to determine whether the US economy is in a slack state. When the slack state is defined as the period of the unemployment rate higher than about 12 percent, the estimated cumulative multipliers are significantly larger during slack periods than non-slack periods and are above unity. Our estimation results emphasize the importance of endogenous sample splitting. Furthermore,  the effect of the fiscal policy may be heterogenous with respect to the level of slackness in the economy, thereby calling for more research in understanding the heterogenous effects of fiscal policy.
Finally, our paper sheds light on the prospect of fiscal policy in response to economic shocks from the current COVID-19 pandemic.

\bibliographystyle{chicago}
\bibliography{LLSS_bib}

\begin{thebibliography}{}

\bibitem[\protect\citeauthoryear{Auerbach and Gorodnichenko}{Auerbach and
  Gorodnichenko}{2012}]{Auerbach12}
Auerbach, A.~J. and Y.~Gorodnichenko (2012).
\newblock Measuring the output responses to fiscal policy.
\newblock {\em American Economic Journal: Economic Policy\/}~{\em 4\/}(2),
  1--27.

\bibitem[\protect\citeauthoryear{Auerbach and Gorodnichenko}{Auerbach and
  Gorodnichenko}{2013a}]{Auerbach13a}
Auerbach, A.~J. and Y.~Gorodnichenko (2013a).
\newblock Fiscal multipliers in recession and expansion.
\newblock In A.~Alesina and F.~Giavazzi (Eds.), {\em Fiscal Policy after the
  Financial Crisis}, pp.\  63--98. University of Chicago Press.

\bibitem[\protect\citeauthoryear{Auerbach and Gorodnichenko}{Auerbach and
  Gorodnichenko}{2013b}]{Auerbach13b}
Auerbach, A.~J. and Y.~Gorodnichenko (2013b).
\newblock Output spillovers from fiscal policy.
\newblock {\em American Economic Review\/}~{\em 103\/}(3), 141--46.

\bibitem[\protect\citeauthoryear{Bai, Chen, Tai-Leung~Chong, and Xin~Wang}{Bai
  et~al.}{2008}]{Bai:2008}
Bai, J., H.~Chen, T.~Tai-Leung~Chong, and S.~Xin~Wang (2008).
\newblock Generic consistency of the break-point estimators under specification
  errors in a multiple-break model.
\newblock {\em Econometrics Journal\/}~{\em 11\/}(2), 287--307.

\bibitem[\protect\citeauthoryear{Blanchard and Perotti}{Blanchard and
  Perotti}{2002}]{BP:02}
Blanchard, O. and R.~Perotti (2002).
\newblock {An Empirical Characterization of the Dynamic Effects of Changes in
  Government Spending and Taxes on Output}.
\newblock {\em Quarterly Journal of Economics\/}~{\em 117\/}(4), 1329--1368.

\bibitem[\protect\citeauthoryear{Hansen}{Hansen}{1996}]{Hansen:1996}
Hansen, B.~E. (1996).
\newblock Inference when a nuisance parameter is not identified under the null
  hypothesis.
\newblock {\em Econometrica\/}~{\em 64\/}(2), 413--430.

\bibitem[\protect\citeauthoryear{Hansen}{Hansen}{2000}]{hansen2000sample}
Hansen, B.~E. (2000).
\newblock Sample splitting and threshold estimation.
\newblock {\em Econometrica\/}~{\em 68\/}(3), 575--603.

\bibitem[\protect\citeauthoryear{Hidalgo, Lee, and Seo}{Hidalgo
  et~al.}{2019}]{hidalgo2019robust}
Hidalgo, J., J.~Lee, and M.~H. Seo (2019).
\newblock Robust inference for threshold regression models.
\newblock {\em Journal of Econometrics\/}~{\em 210\/}(2), 291--309.

\bibitem[\protect\citeauthoryear{Jord{\`a}}{Jord{\`a}}{2005}]{jorda2005estimation}
Jord{\`a}, {\`O}. (2005).
\newblock Estimation and inference of impulse responses by local projections.
\newblock {\em American Economic Review\/}~{\em 95\/}(1), 161--182.

\bibitem[\protect\citeauthoryear{Lee, Liao, Seo, and Shin}{Lee
  et~al.}{2018}]{lee2018factor}
Lee, S., Y.~Liao, M.~H. Seo, and Y.~Shin (2018).
\newblock Factor-driven two-regime regression.
\newblock {\em arXiv preprint\/}.
\newblock \url{https://arxiv.org/abs/1810.11109}.

\bibitem[\protect\citeauthoryear{Newey and West}{Newey and West}{1987}]{NW87}
Newey, W. and K.~West (1987).
\newblock A simple, positive semi-definite, heteroskedasticity and
  autocorrelation consistent covariance matrix.
\newblock {\em Econometrica\/}~{\em 55}, 703--708.

\bibitem[\protect\citeauthoryear{Ramey and Zubairy}{Ramey and
  Zubairy}{2018}]{RZ-2017}
Ramey, V. and S.~Zubairy (2018).
\newblock Government spending multipliers in good times and in bad: Evidence
  from us historical data.
\newblock {\em Journal of Political Economy\/}~{\em 126\/}(2), 850--901.

\bibitem[\protect\citeauthoryear{Ramey}{Ramey}{2011}]{ramey2011identifying}
Ramey, V.~A. (2011).
\newblock Identifying government spending shocks: It's all in the timing.
\newblock {\em Quarterly Journal of Economics\/}~{\em 126\/}(1), 1--50.

\bibitem[\protect\citeauthoryear{Weir}{Weir}{1992}]{weir1992century}
Weir, D.~R. (1992).
\newblock A century of us unemployment, 1890-1990: Revised estimates and
  evidence for stabilization.
\newblock {\em Research in Economic History\/}~{\em 14\/}(1), 301--46.

\end{thebibliography}

\end{document}